\documentclass[leqno,11pt]{amsart}

\usepackage{amsmath,amssymb,amsfonts,amsthm}
\usepackage{graphics}
\usepackage{graphicx}
\usepackage{subfigure}
\usepackage[a4paper,total={6in,8in}]{geometry}
\usepackage{enumerate}

\begin{document}

\title[Pricing Weather Derivatives]{Neural and Time-Series Approaches for Pricing Weather Derivatives: Performance and Regime Adaptation using Satellite Data}

\author[]{Marco Hening Tallarico, Toronto Metropolitan University, Pablo Olivares, Florida International University}


\begin{abstract}
This paper investigates the pricing of weather-derivative (WD) contracts whose underlying variables are temperature or precipitation. For temperature-linked contracts we compare a harmonic-regression/ARMA benchmark with a feed-forward neural-network forecast; in both Toronto and Chicago the network reduces out-of-sample mean-squared error (MSE) and materially shifts the fair value of December strangle contracts relative to the time-series model and to the industry-standard Historic Burn Approach (HBA).

For precipitation we adopt a compound Poisson–Gamma framework, estimating shape and scale parameters by maximum likelihood and a convolutional neural network (CNN) trained on 30‑day rainfall sequences spanning multiple seasons. The CNN adapts Gamma parameters to seasonal regimes. While less precise than MLE estimates, this regime-adaptive approach implicitly captures seasonal dependence, addressing the limitation of static i.i.d. models. At valuation, the approach assumes days are i.i.d. $\Gamma(\hat{\alpha}, \hat{\beta})$ within each seasonal regime and uses a mean-count approximation for analytical tractability, yielding a closed-form strangle price.

Our exploratory analysis of historical rainfall shows clear seasonal heterogeneity: the Gamma shape and scale parameters differ materially from summer to winter, underscoring the need for season-specific precipitation models. This pronounced seasonal divergence indicates that a static, global parameter fit is insufficient and strongly motivates our CNN’s regime-adaptive estimation strategy, which learns distinct $(\alpha,\beta)$ mappings for each season.

All models are trained and back-tested on the 1981–2023 NASA POWER gridded data set for Toronto (Canada) and Chicago (USA). Expected pay-offs are evaluated analytically whenever possible and otherwise by simulation, enabling a like-for-like comparison of forecasting techniques, valuation methods, and their economic consequences.
\end{abstract}


\maketitle
\pagestyle{plain}

\newpage
\section{Introduction}

Extreme and increasingly volatile weather exposes energy companies, farmers,
airlines, and municipalities to earnings swings that traditional insurance
cannot easily absorb. A global market for weather-derivative (WD) contracts
has therefore emerged, with notional volume now estimated in the tens of
billions of U.S.\ dollars each year.\footnote{Weather Risk Management
Association annual survey, 2024.} Accurate and transparent pricing of those
contracts is thus commercially and regulatorily important.

Three families of pricing methods dominate the literature:

\begin{enumerate}
\item \textbf{Historic Burn Approach (HBA).}  
      The mean of discounted past pay-offs, often uplifted by a risk premium.
      Popular with practitioners, but ignores structural climate change.

\item \textbf{Parametric time-series models.}  
      Harmonic regression, ARMA-type dynamics, or stochastic-differential
      models combined with Fourier inversion \cite{alexa,jewson}. These assume
      linear dependence and require long homogeneous station records.

\item \textbf{Machine-learning forecasts.}  
      Neural networks can exploit nonlinear patterns in climate data, but
      their use in WD pricing—especially for precipitation—is scant.
\end{enumerate}

Existing approaches also make two restrictive assumptions:
\begin{enumerate}[(i)]
  \item daily precipitation amounts are independent and identically distributed (i.i.d.);
  \item local station data are available and error‑free.
\end{enumerate}
Both are violated in practice.

\bigskip

\noindent\emph{How our method improves on these limitations:}
\begin{itemize}
  \item \emph{Regime‑shift flexibility in temperature modeling.}
    Whereas HBA, parametric time‑series models, and standard ML forecasts all
    treat temperature as coming from a single, homogeneous statistical regime, our CNN dynamically learns and adapts parameter estimates
    $(\alpha, \beta)$ to seasonal and regime changes in the temperature process.
\end{itemize}

Earlier work has tackled the risk-neutral layer of temperature markets by embedding a time-changed Lévy driver into an OU dynamics, which yields closed-form Esscher prices and complements the purely actuarial valuation we adopt here \cite{Olivares_2}. On the statistical side, Zapranis and Alexandridis used a feed-forward network to model the seasonal residual variance of an OU temperature process, foreshadowing our own use of neural networks to capture heteroskedasticity in both temperature and precipitation indices \cite{Achilleas}. Existing approaches also make two restrictive assumptions. 

This work extends our earlier SARIMA analysis of Toronto temperatures
\cite{tallarico2023SARIMA}, showing that seasonal-ARIMA forecasts can feed
directly into the strangle-pricing formula and yield competitive benchmark
prices for WD contracts.

\medskip
\noindent
\textbf{Contributions.}  
We address those gaps as follows.

\begin{itemize}
  \item \emph{Satellite‐derived climate data.}
    We are the first to employ the 1981--2023 NASA POWER gridded data set for
    weather‑derivative pricing, providing uniform coverage where gauges are sparse.
  
  \item \emph{Neural estimation for regime‑adaptive parameter estimation.}
    \begin{enumerate}[(a)]
      \item \textbf{Temperature.}
        A lag‑augmented feed‑forward network is benchmarked against a harmonic‑regression/ARMA model.
        By conditioning on recent values, the model flexibly adapts to short‑term autocorrelations
        while retaining a simple residual noise assumption.
      \item \textbf{Precipitation.}
        We reshape each 30‑day rainfall series into a one‑dimensional “image” and train a CNN to
        learn wet--dry patterns and seasonal shifts. The network’s outputs—season‑specific estimates of
        the Gamma shape and scale—are then plugged into the Gamma‑sum formula to obtain a
        closed‑form strangle price. The CNN’s benefit is solely in adapting parameters $\alpha$ and $\beta$
        across regimes.
    \end{enumerate}
\end{itemize}

Empirically, our feed-forward network reduces one-month-ahead temperature-forecast MSE relative to harmonic/ARMA, while our CNN rainfall model matches the i.i.d.–Gamma approach in pricing accuracy and simultaneously captures the wet–dry autocorrelation that the Gamma baseline omits.

Although our models are task-specific, Chen et al.\ \cite{chen2023foundation} point to the growing role of foundation models in climate and geospatial tasks, suggesting future extensions of our framework using pretrained, scalable architectures. Unlike \cite{chen2023foundation}, which discusses generalized climate understanding, our work centers on narrow but actionable financial use cases, specifically strangle-option pricing for December temperature and precipitation exposures in two North American cities.\\

\textbf{The remainder of the paper is organised as follows.}\\[1ex]
\textbf{Section~2} fits and forecasts temperature via harmonic regression/ARMA;\\
\textbf{Section~3} presents the neural forecast;\\
\textbf{Section~4} estimates precipitation parameters by MLE and CNN;\\
\textbf{Section~5} prices WD strangles under each model;\\
\textbf{Section~6} concludes.

\section{Temperature data and harmonic-ARMA modeling}

Let $(T_t)_{t=0,1,...,n}$ be a sequence of random variables representing the temperature at certain region on day $t$. Typically, daily temperature  data consist of an average of the maximum and the minimum values during the day. We notice the presence of a seasonal component as empirical data show.\\
The first model for temperature  combines  a harmonic regression with an ARMA(2,3) model. We also investigated the addition of a GARCH(1,1) model, however the latter does not show any improvement in the fitting for our particular data.\\
The forecast for daily temperature is done  for the month of December 2023. The data  have been obtained from the NASA Earth Science/Applied Science Program (website:https://power.larc.nasa.gov). Observations are recorded through satellite. As such, they could be less accurate as measuring temperature data from satellites requires statistical models to process the data gathered.  The time period covered by the data ranges from 01/01/1981- 12/31/2023. Specific locations  in Toronto and Chicago data collection are shown in Table \ref{tab:location}.

\begin{table}[hbt!]
\centering
\begin{tabular}{|c|c|c|}
\hline
Location & Lat. & Long. \\ \hline
Toronto  & 43.6523 & -79.3839 \\
Chicago &41.4047 &-89.6420 \\\hline
\end{tabular}
\caption{Latitude and longitude locations of the data from NASA's POWER project}
\label{tab:location}
\end{table}

\begin{figure}[hbt!]
\begin{center}
\subfigure[]{
\resizebox*{7cm}{!}{\includegraphics{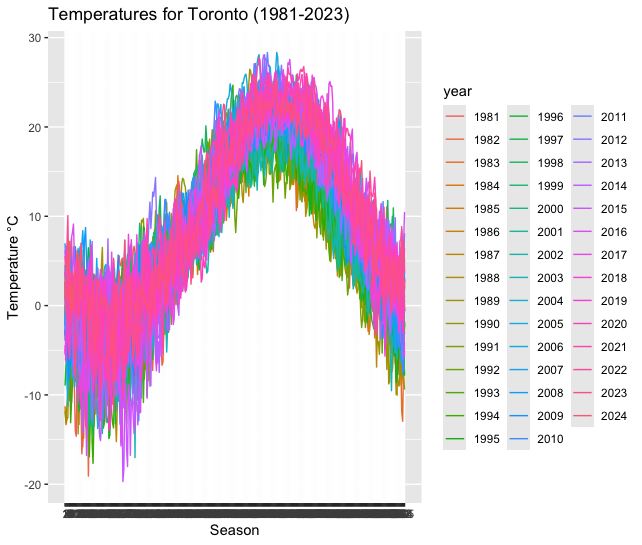}}}
\subfigure[]{
\resizebox*{7cm}{!}{\includegraphics{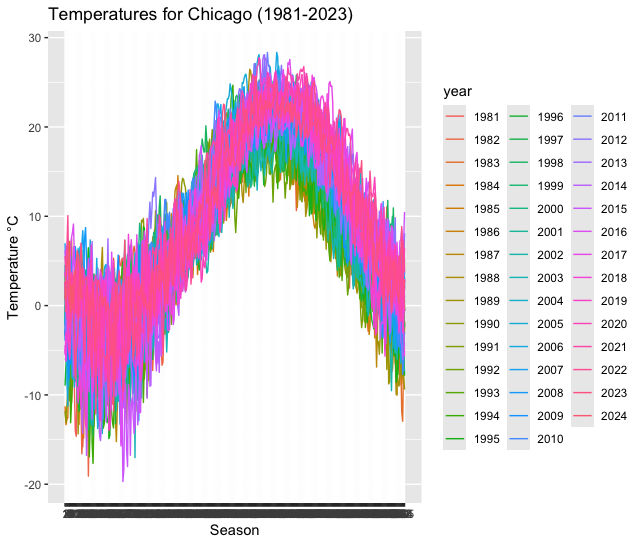}}}
 \caption{Left: Daily temperature data for Toronto from 1981-2023. Right: Daily temperature data for Chicago from 1981-2023}\label{fig:temp}
\end{center}
\end{figure}

In figures \ref{fig:temp}(a) and \ref{fig:temp}(b) the values of temperatures in Toronto and Chicago cities, measured in Celsius degrees, are respectively shown. Each curve represents a particular year. Seasonal effects of temperatures on both cities can be observed. However, seasonality is less clearly seen in the graph for precipitation. Table \ref{tab:stats} complement the information with some statistics of the temperature and precipitation data.  \\
Notice that the negative  value of the kurtosis in temperatures  indicates the presence of tails lighter than those of the normal distribution, while the precipitation exhibits a large kurtosis in both cities, even  larger in Chicago. Also, the data is right skewed. The average number of rainy days in December for Toronto and Chicago are respectively $29.79$ and $23.14$.\\

\begin{table}[htb!]
\centering
\begin{tabular}{|c|c|c|c|c|c|c|}
\hline
Location  &  Minimum &  Maximum  & 	Mean & Std Dev  &  Skewness   &  Kurtosis \\ \hline 	 		
Toronto Temp.  & -19.7  & 28.35   &  8.37017 &  9.254303   & -0.053197  & -0.949363 \\
Toronto precip.  & 0.01  & 41.59   &  2.04 & 3.34   & 3.36  &  16.11 \\  \hline
Chicago Temp.  & -31  &  33.28  & 9.690668  &  11.42069   & -0.3264819  &  -0.777556 \\ 
Chicago precip.  & 0.01   & 75.93  & 3.47   &6.41    & 3.59  &  19  \\ \hline
\end{tabular}
\caption{Statistical summary of temperature and precipitation time series in Toronto and Chicago}
\label{tab:stats}
\end{table}

\begin{figure}[hbt!]
\begin{center}
\subfigure[]{
\resizebox*{7cm}{!}{\includegraphics{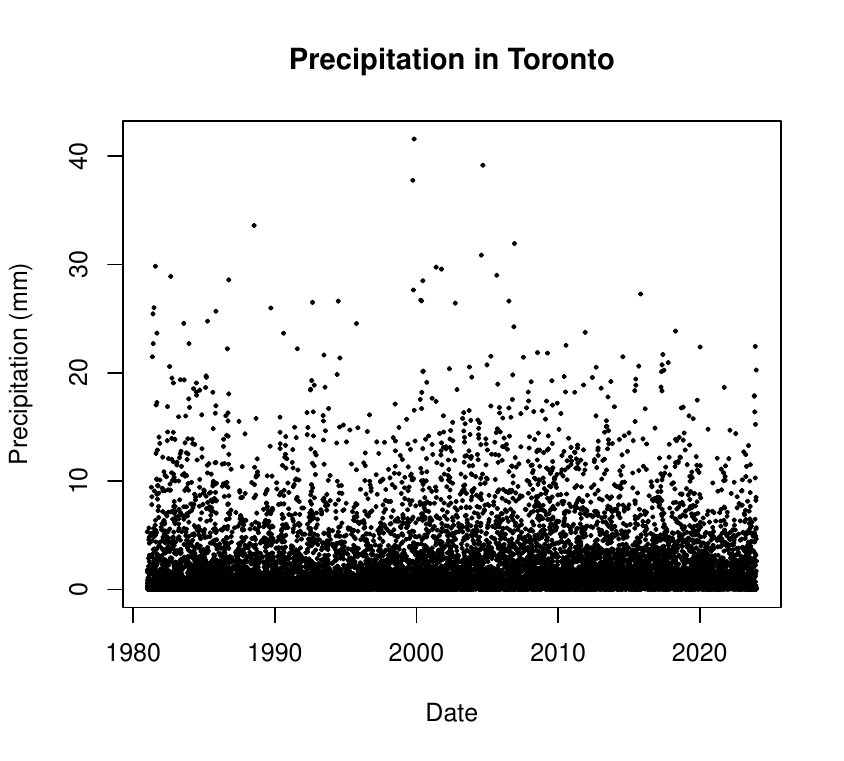}}}
\subfigure[]{
\resizebox*{7cm}{!}{\includegraphics{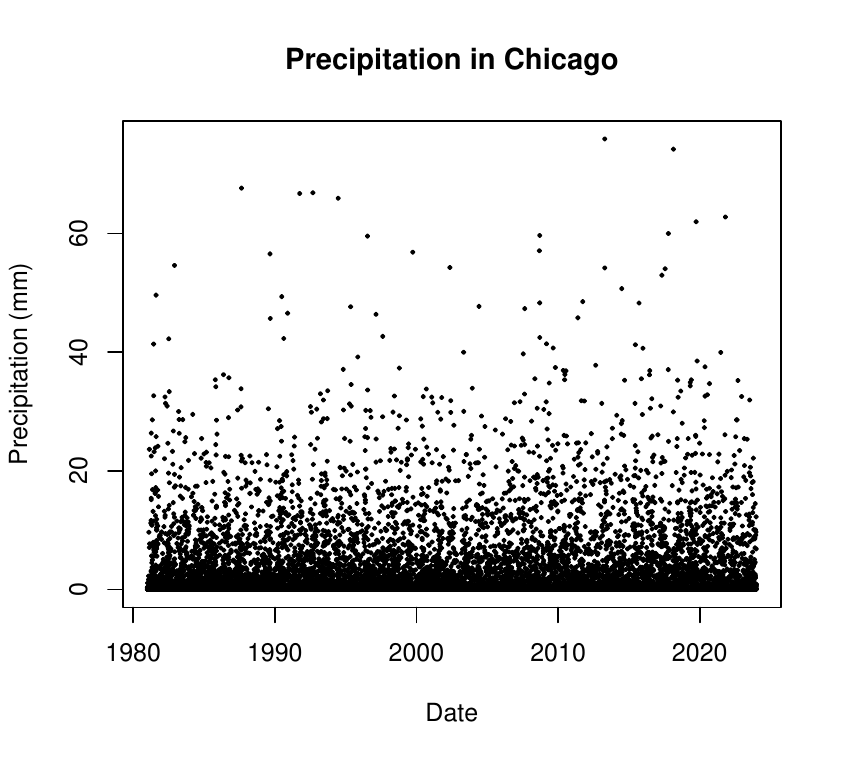}}}
 \caption{ Daily precipitation data for Toronto and Chicago cities from 1981-2023}\label{fig:TO_precip} 
\end{center}
\end{figure}

In Figures \ref{fig:TO_precip}(a) and \ref{fig:TO_precip}(b)  daily precipitation data in Toronto and Chicago from 1981-2023 are respectively shown.

Results about the fitting  a harmonic regression to data are given in the Table \ref{tab:harmonic estimators}. Notice that all coefficients are significantly different from zero, as reflected in the  p-value.

\begin{table}[hbt!]
        \centering
        \begin{tabular}{|c|c|c|c|c|}
        \hline
        Toronto &   &   & & \\ \hline
        Parameter & Estimate     & SE       & t-Stat  & p-Value   \\ 
        $\beta_0$ & 6.904  & 6.613e-02  & 104.39  & \textless{}2e-16 \\
        $\beta_1$& 8.665e-05   & 7.482e-06  & 11.58   & \textless{}2e-16 \\
        $\beta_2$ & -6.809 & 4.673e-02  & -145.69 & \textless{}2e-16 \\
        $\beta_3$ & -1.244e+01 & 4.679e-02  & -265.90 & \textless{}2e-16 \\ \hline
        Chicago &   &   &  & \\ \hline
Parameter & Estimate     & SE       & t-Stat  & p-Value   \\ 
        $\beta_0$ & 6.904  & 6.613e-02  & 104.39  & \textless{}2e-16 \\
        $\beta_1$& 8.665e-05   & 7.482e-06  & 11.58   & \textless{}2e-16 \\
        $\beta_2$ & -6.809e+00 & 4.673e-02  & -145.69 & \textless{}2e-16 \\
        $\beta_3$ & -1.244e+01 & 4.679e-02  & -265.90 & \textless{}2e-16 \\ \hline
        \end{tabular}
        \caption{ Coefficients and statistics of harmonic regression for daily temperature data in Toronto and Chicago.} \label{tab:harmonic estimators}
\end{table}

\begin{figure}[hbt!]
\begin{center}
\subfigure[]{
\resizebox*{7cm}{!}{\includegraphics{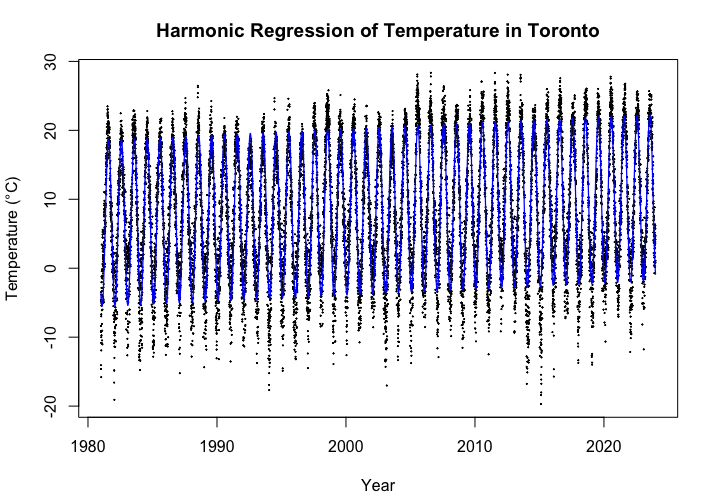}}}
\subfigure[]{
\resizebox*{7cm}{!}{\includegraphics{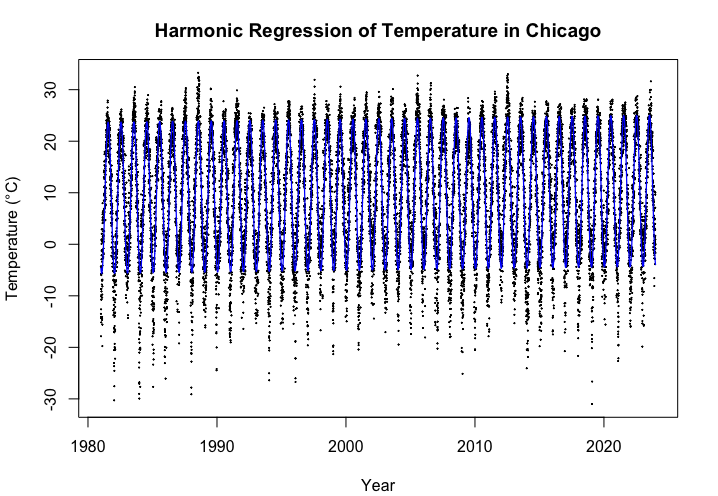}}}
 \caption{ Harmonic regression of daily temperature time series in Toronto and Chicago from 1981 - 2023. The blue curve represents the regression and the black dots are the actual temperatures.}\label{fig:hrm_regress} 
\end{center}
\end{figure}

\begin{figure}[hbt!]
\begin{center}
\subfigure[]{
\resizebox*{7cm}{!}{\includegraphics{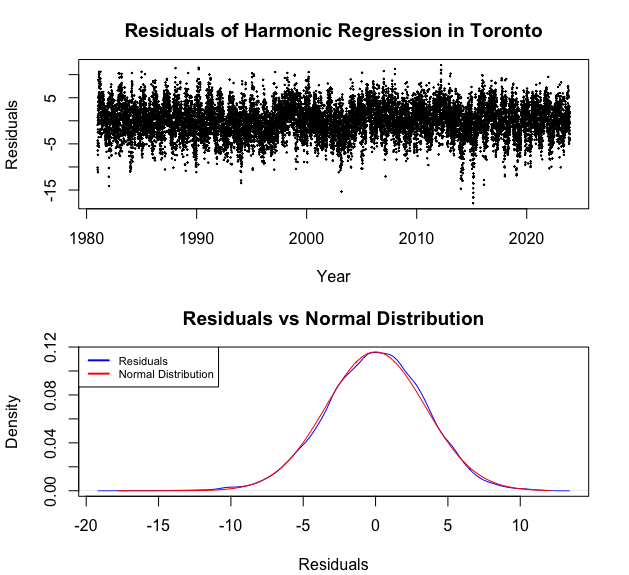}}}
\subfigure[]{
\resizebox*{7cm}{!}{\includegraphics{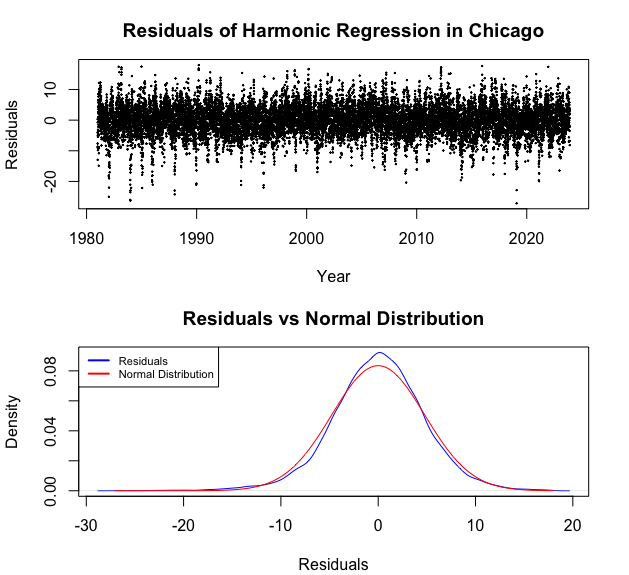}}}
 \caption{The top figure shows the scatter plot for the residuals of the harmonic regression and the bottom shows the density plot of these residuals compared with the density plot of the normal distribution }\label{fig:hrm_res} 
\end{center}
\end{figure}

  In figures \ref{fig:hrm_regress}(a) and \ref{fig:hrm_regress}(b) harmonic regression fit for  daily temperature time-series in Toronto and Chicago are represented. The blue curve represents the regression and the black dots are the actual temperatures. On the other hand, figures \ref{fig:hrm_res}(a) and \ref{fig:hrm_res}(b) show the residual values in both fittings. It can be appreciated that the residuals from the harmonic regression do not resemble white noise. By performing an Augmented Dicky-Fuller test on the residuals we see that they are stationary. This fact suggests an ARIMA model can be used for the residuals. The top figure shows the scatter plot for the residuals of the harmonic regression and the bottom shows the density plot of these residuals compared with the density plot of the normal distribution.

A Kolmogorov-Smirnov test for the temperature data in both cities rejects  t-student, stable and inverse Gaussian distributions. In both cases the test fails to reject the normality. On the other hand, auto-correlation and partial auto-correlation  in the residuals could be observed.\\
The values $p$ and $q$ in the  ARMA model are selected to minimize the Akaike information criterion (AIC) and Bayesian information criterion (BIC). We conducted  testing on models with autoregressive(AR) and moving average (MA) components up to order five. The optimal values of the order according to the Akaike criterion are $p=2$ and $q=3$ for both, Toronto and Chicago cities.\\
We  also look at the largest significant lags in the autocorrelation (ACF) and partial autocorrelation  (PACF) functions  to further justify the choice of an ARMA(2,3) model. The ACF of the residuals contains several statistically significant lags. From the PACF function we see that there are two major spikes, as such we should expect the ARMA model to have an autoregressive component of two our model by observation be ARMA(p=2, q=3). \\
The estimation of the coefficients is done  by a maximum likelihood approach. Their values and  their corresponding standard errors can be seen in Table \ref{tab:TO_ARMA}.

\begin{table}[hbt!]
\centering
\begin{tabular}{|c|c|c|c|c|c|}
\hline
Toronto &$\phi_1$ &$\phi_2$ & $\theta_1$ & $\theta_2$ & $\theta_3$  \\ \hline
Estimate &1.5426  &-0.5511   &-0.7387    &-0.2842  &0.0947  \\ \hline
S.E &0.0357  &0.0343 &0.0375 &0.0120 &0.0213 \\ \hline
Chicago &$\phi_1$ &$\phi_2$ & $\theta_1$ & $\theta_2$ & $\theta_3$  \\ \hline
Estimate &1.5256 &-0.5475  &-0.6036  &-0.3226 & 0.0443 \\ \hline
S.E &  0.0384   &0.0332    &0.0396  &0.0097 &0.0173\\ \hline
\end{tabular}
\caption{ARMA(2,3) coefficients and their standard error of coefficients in Toronto and Chicago}\label{tab:TO_ARMA}
\end{table}

\begin{figure}[htb!]
\begin{center}
\subfigure[]{
\resizebox*{7cm}{!}{\includegraphics{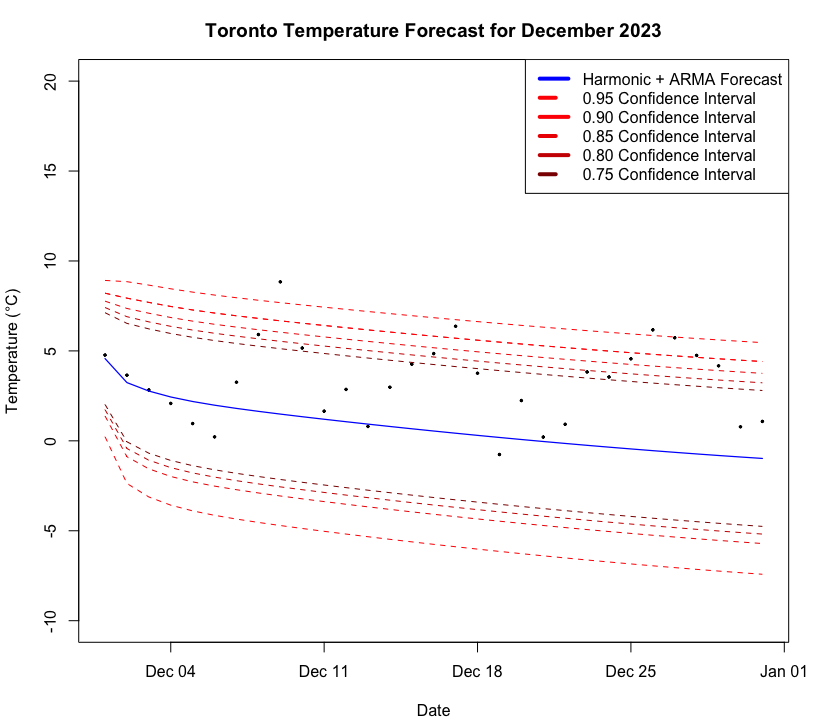}}}
\subfigure[]{
\resizebox*{7cm}{!}{\includegraphics{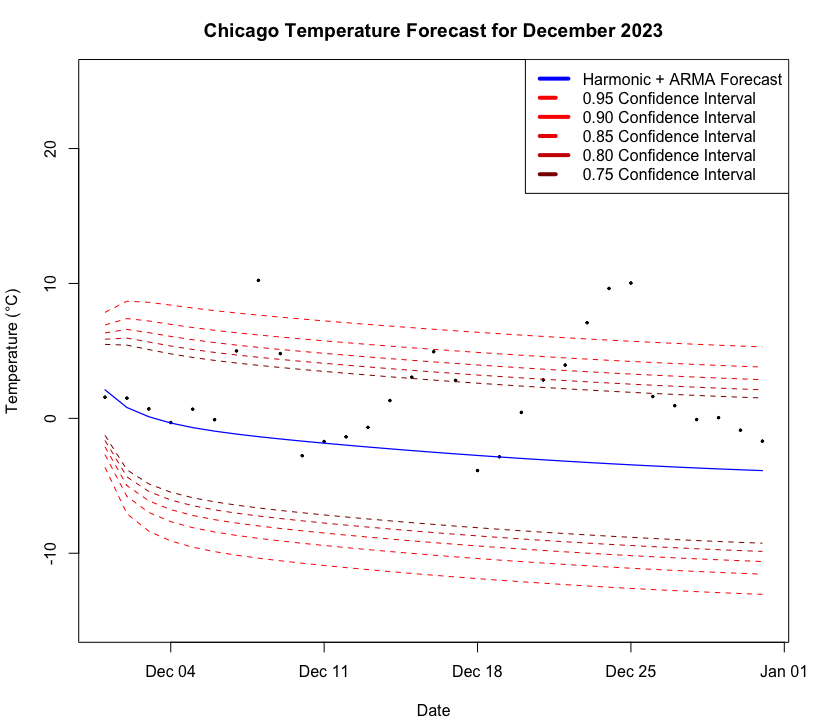}}}
 \caption{Daily temperature forecast from harmonic regression and ARMA model for Toronto and Chicago, December 2023.}\label{fig:hrm_arma_fore} 
\end{center}
\end{figure}

In Figures \ref{fig:hrm_arma_fore}(a) and  \ref{fig:hrm_arma_fore}(b) the forecast of temperatures for the month of December 2023 are shown. The blue curve at the center is the daily forecast with the red curves are the confidence limits with confidence levels ranging from 75\% and 95\%.

The time series analysis has been carried out using R studio.

\section{Forecasting temperatures: Neural-network approach}

Recent surveys, such as \cite{zhang2024mlweather}, provide comprehensive overviews of machine learning techniques for weather forecasting, highlighting both classical methods (e.g., ARIMA, SVM) and deep learning architectures (e.g., CNNs, RNNs). This supports our use of feed-forward and convolutional neural networks as modern alternatives to traditional time-series models in WD pricing. However, unlike \cite{zhang2024mlweather}, which emphasizes generic forecasting accuracy, our work explicitly ties forecasting output to the financial valuation of derivative contracts, providing an end-to-end pricing methodology.

Shi et al.\ \cite{shi2025deepweather} systematically classify deep learning models for weather prediction and describe training paradigms applicable to our feed-forward and CNN approaches. Their taxonomy underscores the relevance of neural methods in predictive meteorology and informs our model selection. In contrast, our contribution lies in directly integrating these predictions into an analytical pricing framework for weather derivatives, rather than building general-purpose or multi-scale weather forecasting systems.

\begin{figure}[hbt!]
\begin{center}
\subfigure[]{
\resizebox*{7cm}{!}{\includegraphics{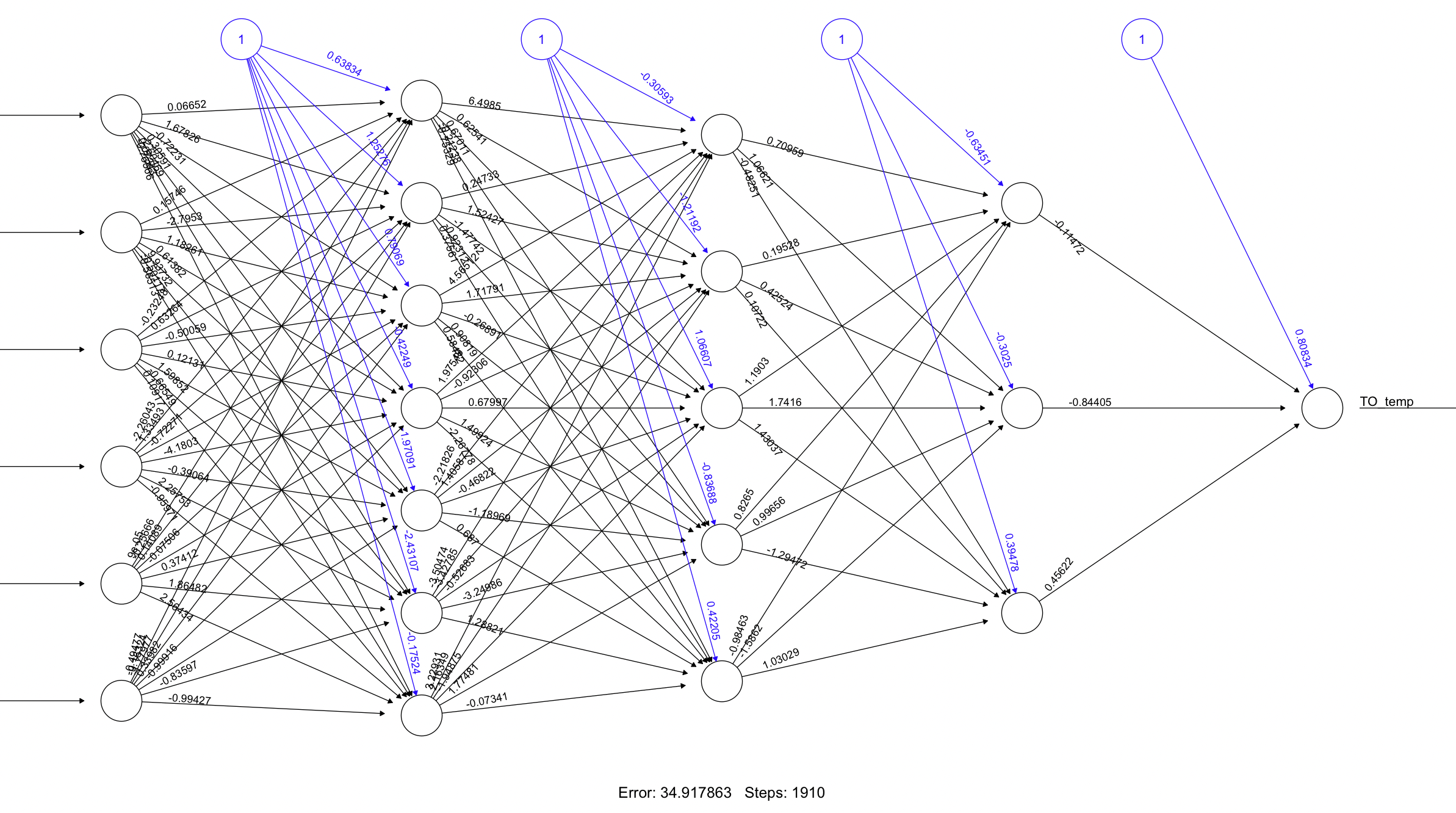}}}
\subfigure[]{
\resizebox*{7cm}{!}{\includegraphics{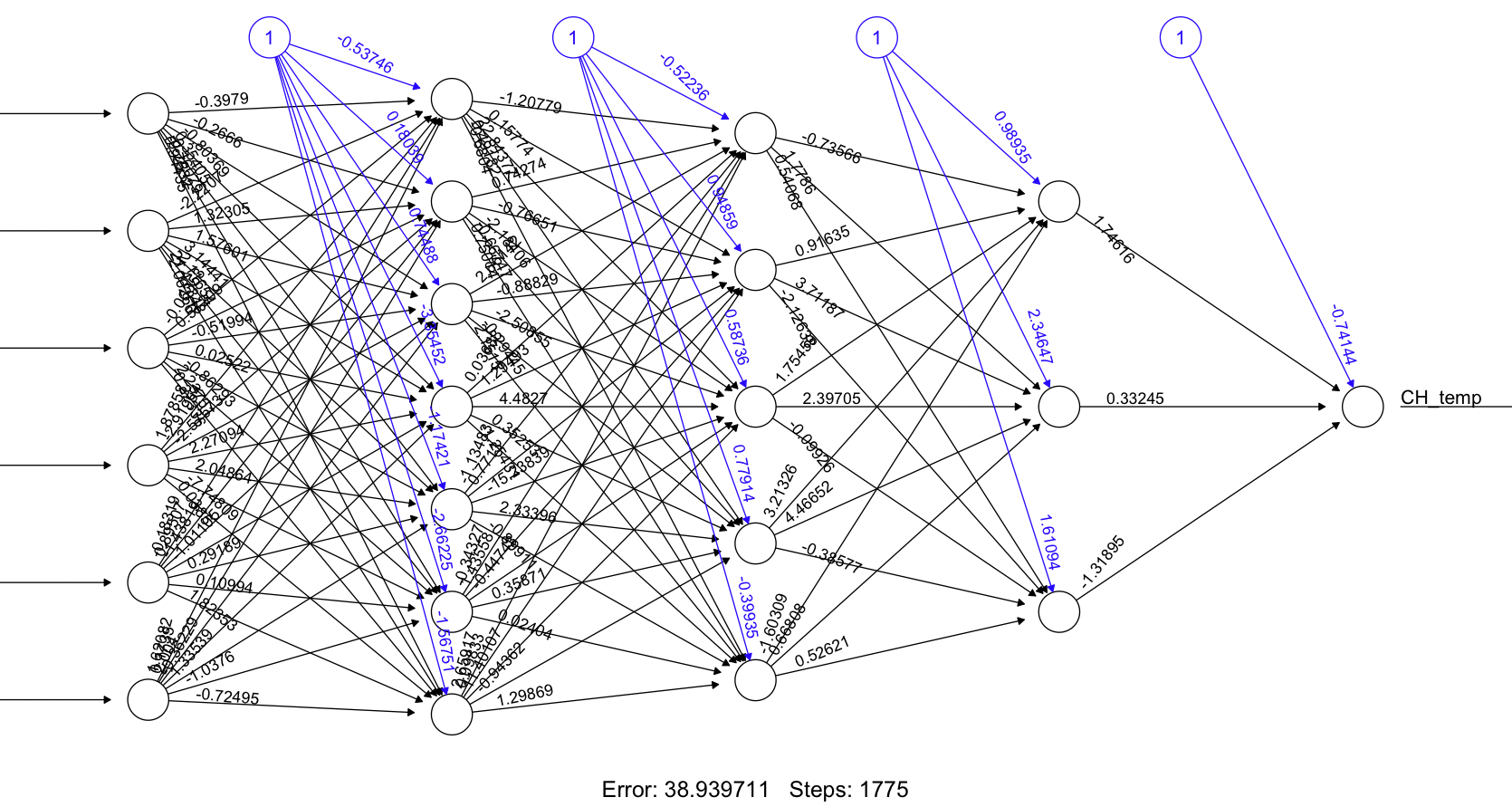}}}
 \caption{Neural networks used for Toronto(left) and Chicago(right) temperature forecasts.}\label{fig:T_TO_NN} 
\end{center}
\end{figure}

\begin{figure}[hbt!]
\begin{center}
\subfigure[]{
\resizebox*{7cm}{!}{\includegraphics{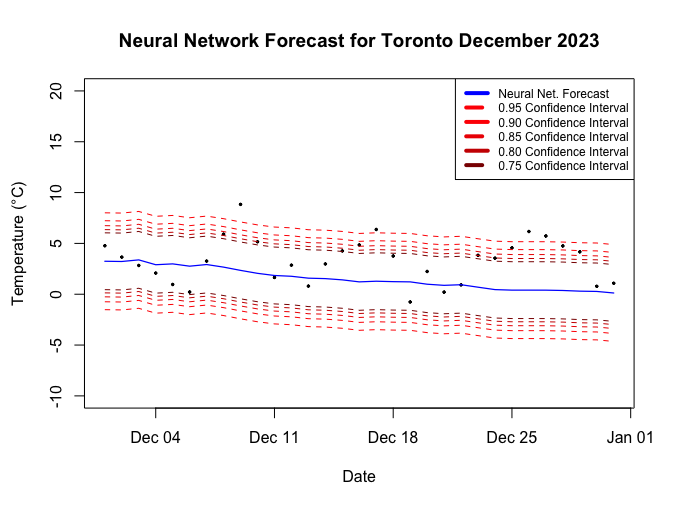}}}
\subfigure[]{
\resizebox*{7cm}{!}{\includegraphics{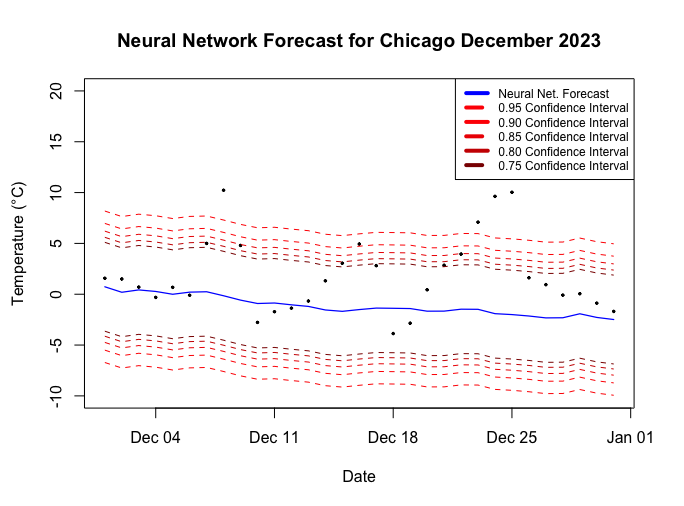}}}
 \caption{Daily temperature forecast from neural network for Toronto and Chicago, December 2023.}\label{fig:T_TO_NN_fore} 
\end{center}
\end{figure}

\begin{figure}[hbt!]
\begin{center}
\subfigure[]{
\resizebox*{7cm}{!}{\includegraphics{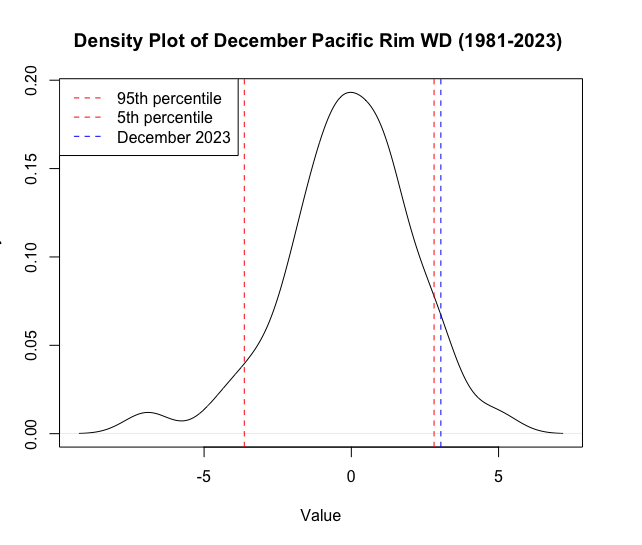}}}
\subfigure[]{
\resizebox*{7cm}{!}{\includegraphics{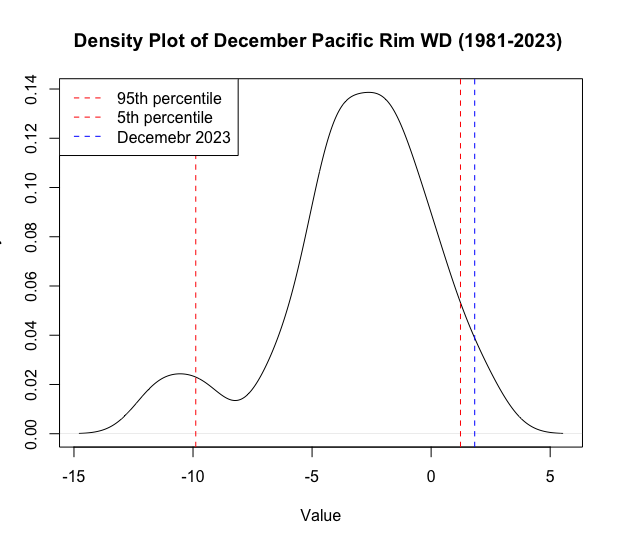}}}
 \caption{Density p.d.f. for December Pacific Rim indices (1981-2023) in Toronto and Chicago cities}\label{fig:T_TO_PRIM_density} 
\end{center}
\end{figure}

We  consider a neural network to train and forecast temperature data. We will benchmark this  neural network model to two more traditional models considered in the previous section.\\
 The neural networks into consideration have three hidden layers with (7,5,3) neurons in each layer.
 The six inputs we use in the neural network for Toronto and Chicago are the following: year, month, day, temperature one year ago, temperature two years ago, and the data from the harmonic regression. To keep the dimensional of the data low, we chose to encode year, month, and day data using label encoding.\\
 We chose this method of encoding categorical variables for two main reasons. 
 First, year, month, and day have a natural chronological order. Second, we want to keep the dimensionality of the data to be low. It is important to note that one-hot encoding might be a better choice to represent the data, but  at the expense of increased dimensionality. The use of this encoding is common  in machine learning when categorical data needs to be represented numerically, see \cite{Time_series_R, ML_python}.\\ 
 We also use the temperature of the previous two years as an input to the network because the harmonic regression is not able to model all the seasonality present in the data. By adding the values of the harmonic regression and the previous two years of daily average temperature as inputs we intend a better capture these seasonal dynamics. \\
 Because we take the temperature from two years ago as an input, the neural network is trained on the remaining 40 years of historical data. We apply  \textit{min-max normalization} on the features and log-normalization on target variable (DAT).\\
This normalization is done in order to speed up the training time, and the convergence of our model, see \cite{batch}. A \textit{Resistant Back Propagation Algorithm} is used to find the weights in the models.\\
 We arrive at the architecture exposed above by measuring the Mean Square Error (MSE) of different neural networks on a validation set that is withheld from the training  and testing sets. Neural networks up to three hidden layers and ten neurons per layer were tested. Using a validation set to find the best architecture for a neural network is best practice. Performing this testing on the testing set could lead to data-leakage and an over fit model. See \cite{StatLearn_R}. While the number of neurons is the same in the neural network that forecasts temperature in Chicago and Toronto, all the weights in each neural network are different because each neural network was trained on a different set of data. A visualization of the neural network can be seen in Figure \ref{fig:T_TO_NN}.\\ 
 The forecast of the temperatures for Toronto in December 2023 can be seen in Figure \ref{fig:T_TO_NN_fore}. We can see that both forecasts still underestimate the temperature for December 2023. There are multiple values in our forecast that pass the upper 95 \% confidence interval. A major reason of this underestimation  is that  this particular month in 2023 was unusually hot when compared to other months in December during past years.\\
 In Figure \ref{fig:T_TO_PRIM_density} we can see the p.d.f. plot of average temperatures in December from 1981-2023. In this graph the red lines represent the 0.95 confidence interval for temperature in December and the blue line represents the observed temperature in December 2023. From the graph we can see that December 2023 seems an outlier month with unusually high temperatures.

\begin{figure}[hbt!]
\begin{center}
\subfigure[]{
\resizebox*{7cm}{!}{\includegraphics{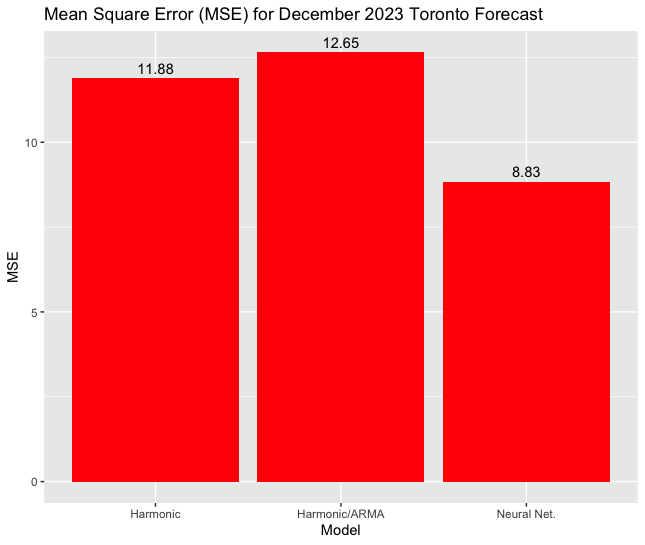}}}
\subfigure[]{
\resizebox*{7cm}{!}{\includegraphics{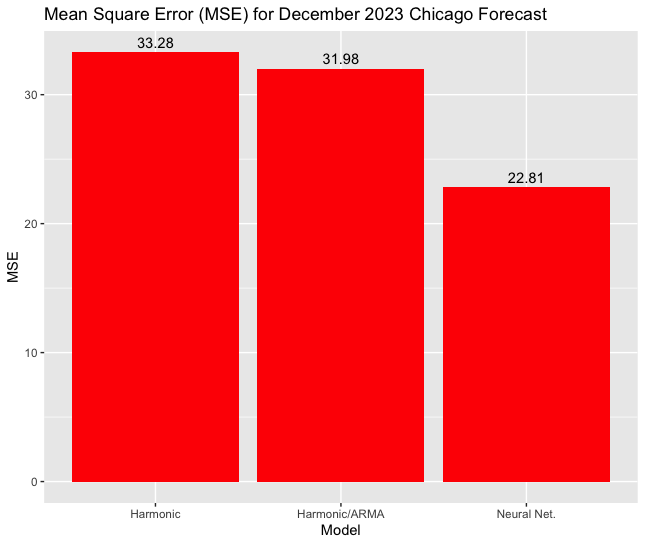}}}
 \caption{Mean Square Error for the harmonic regression, harmonic regression and ARMA model, and neural network for Toronto and Chicago, December 2023 forecast}\label{fig::T_TO_MSE} 
\end{center}
\end{figure}

The underestimation, due to unusually high temperatures for December 2023 is also present in the Chicago forecast. As can be seen in the p.d.f. plot for average temperature in December in Figure \ref{fig:T_TO_PRIM_density}, the average temperature for December 2023 seems to be an outlier month.  One of the possible reasons for these usually high temperatures in 2023 was is the presence of the \textbf{El Ni\~{n}o phenomenon}.  We believe that the neural network model provides a better prediction than the Harmonic and Harmonic/ARMA models because the former is better able to handle the non-linear patterns in the data.  The ARMA model assumes linear relationships in the data which might not be a good assumption when dealing with DAT data. The neural network can identify crucial patterns in  raw data with less manual feature engineering which  saves time and effort. On the other hand over engineering the input data into a neural network model could lead to poor forecasts. The  MSE for the harmonic regression, harmonic regression plus an ARMA model, and the neural network for Toronto and Chicago, December 2023 forecasts are shown in Figure \ref{fig::T_TO_MSE}. It can be noticed that the MSE is smaller when a neural network approach is considered, compared with the one based on a harmonic regression model and a harmonic model combined with an ARMA(2,3) model.

\section{Precipitation model and estimation analysis}
To promote transparency in neural weather models, \cite{yang2024interpretable} survey interpretability techniques like SHAP and Grad-CAM, which are especially pertinent for the adoption of ML-based pricing in regulatory or financial settings. Our focus, however, is not on interpretability but rather on improving pricing accuracy and computational efficiency by removing the i.i.d.\ assumption in precipitation modeling. However, much could be done in this aspect to find the optimal features to forecast the underlying climate phenomena.

\subsection{Compound Poisson–Gamma specification}

\paragraph{}
Let \(N\) be the (random) number of wet days in the \(n\)-day window and \(R_k\) the precipitation on each wet day.  We assume
\[
N\sim\mathrm{Pois}(\lambda),\quad
R_k\mid N\;\text{i.i.d.}\;\Gamma(\alpha,\beta),
\]
and approximate \(N\) by its mean \(n=\mathbb{E}[N]\). In this case n would be the average days of wet days in December. Then
\[
\xi_R=\frac1n\sum_{k=1}^nR_k\sim\Gamma\bigl(n\alpha,\tfrac\beta n\bigr).
\]

\emph{Remark.} This replaces the full compound–Poisson–Gamma mixture by a single Gamma law at its mean count \(n\), thereby yielding a closed-form density but ignoring the variability of \(N\).  If desired, one can quantify the approximation error by comparing moments of the exact mixture with those of its Gamma approximation.

\subsection{Parameter estimation}

\paragraph{\textit{(i) Maximum likelihood.}}
For each city and for each season we obtain $(\hat\alpha,\hat\beta)$ by
numerically maximising the log-likelihood of the wet-day observations.
Standard errors and 95 \% confidence intervals are computed from the
observed-information matrix; results appear in
Table~\ref{tab:gamma_parameters}.

\paragraph{\textit{(ii) Convolutional neural network (CNN).}}
To relax the i.i.d.\ assumption we encode every December (31-day) rainfall
record as a one-row image whose pixel intensity is the day’s precipitation.
A 1-D CNN with two convolution–max-pool blocks and two dense layers is
trained to regress synthetic Gamma samples onto their generating
$(\alpha,\beta)$ values.  
The network’s architecture is summarised below; Table~\ref{tab:gamma_parameters_CNN}
shows the resulting out-of-sample parameter estimates.

\begin{align*}
\textbf{CNN Architecture:}\quad
&\text{Conv1D (32 filters, kernel 3) -- ReLU} \\
&\text{BatchNorm -- MaxPool1D} \\
&\text{Conv1D (32, 3) -- ReLU} \\
&\text{BatchNorm -- MaxPool1D} \\
&\text{Flatten} \\
&\text{Dense(32) -- ReLU -- BatchNorm -- Dropout(0.6)} \\
&\text{Dense(32) -- ReLU -- BatchNorm -- Dropout(0.6)} \\
&\text{Dense(2) (linear output for $\alpha,\beta$)}
\end{align*}

\begin{figure}[htb]
\centering
\subfigure[]{
   \includegraphics[width=0.48\textwidth]{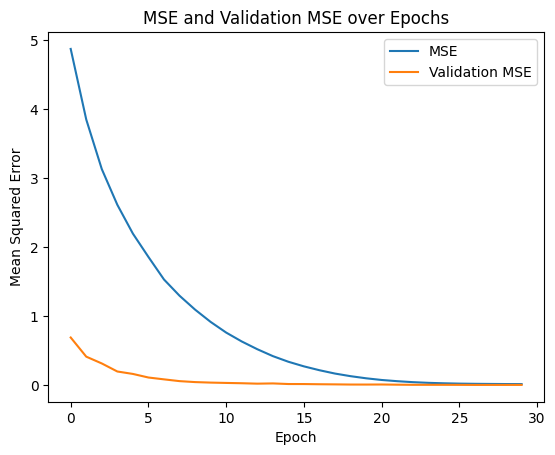}}
\hfill
\subfigure[]{
   \includegraphics[width=0.48\textwidth]{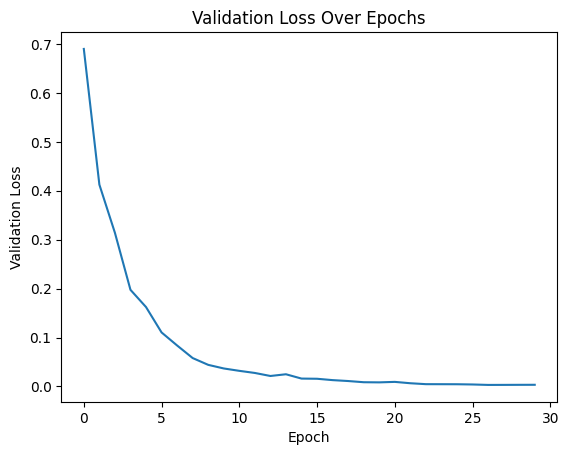}}
\caption{CNN training history: (a) mean-squared error, (b) validation loss.}
\label{fig:cnnmse}
\end{figure}

Training used 100 000 synthetic images with
$\alpha,\beta\!\sim\!\mathrm{Unif}(0.01,5)$ and the Adam optimiser
($\eta\!=\!10^{-3}$).  
Figure~\ref{fig:cnnmse} plots the training and validation MSE, which plateaus after epoch 25, indicating no over-fitting.

\begin{itemize}
  \item \textbf{Relaxed identical distribution:} Unlike MLE (one fixed \(\alpha,\beta\)), the CNN is trained on many \((\alpha,\beta)\) pairs, so it learns across regimes.
  \item \textbf{Preserved independence:} Within each image all days remain independent draws, and in pricing we still assume i.i.d.\ \(\Gamma(\hat\alpha,\hat\beta)\).
\end{itemize}

\begin{table}[htb]
\centering
\caption{Gamma‐parameter estimates via maximum likelihood.}
\label{tab:gamma_parameters}
\begin{tabular}{|l|c|c|c|c|c|c|}
\hline
\textbf{Season} &
$\hat\alpha$ & $\hat\beta$ &
$\mathrm{SE}(\hat\alpha)$ & $\mathrm{SE}(\hat\beta)$ &
95 \% CI$(\hat\alpha)$ & 95 \% CI$(\hat\beta)$ \\
\hline
Toronto (all) & 0.516 & 0.253 & 0.0050 & 0.0038 & [0.506, 0.526] & [0.245, 0.260] \\
Toronto Summer & 0.516 & 0.220 & 0.0101 & 0.0067 & [0.497, 0.536] & [0.207, 0.233] \\
Toronto Winter & 0.580 & 0.377 & 0.0113 & 0.0110 & [0.558, 0.602] & [0.356, 0.399] \\
Toronto Spring & 0.493 & 0.245 & 0.0096 & 0.0075 & [0.474, 0.512] & [0.230, 0.260] \\
Toronto Fall   & 0.504 & 0.221 & 0.0098 & 0.0067 & [0.484, 0.523] & [0.208, 0.235] \\
\hline
Chicago (all)  & 0.371 & 0.107 & 0.0039 & 0.0020 & [0.363, 0.378] & [0.103, 0.110] \\
Chicago Summer & 0.413 & 0.097 & 0.0084 & 0.0034 & [0.397, 0.430] & [0.091, 0.104] \\
Chicago Winter & 0.354 & 0.166 & 0.0076 & 0.0064 & [0.339, 0.369] & [0.154, 0.179] \\
Chicago Spring & 0.391 & 0.104 & 0.0082 & 0.0037 & [0.375, 0.407] & [0.097, 0.111] \\
Chicago Fall   & 0.349 & 0.096 & 0.0076 & 0.0037 & [0.334, 0.364] & [0.088, 0.103] \\
\hline
\end{tabular}
\end{table}

\begin{table}[htb]
\centering
\caption{Gamma‐parameter estimates via CNN.}
\label{tab:gamma_parameters_CNN}
\begin{tabular}{|l|c|c|c|c|c|c|}
\hline
\textbf{Season} &
$\hat\alpha$ & $\hat\beta$ &
$\mathrm{SE}(\hat\alpha)$ & $\mathrm{SE}(\hat\beta)$ &
95 \% CI$(\hat\alpha)$ & 95 \% CI$(\hat\beta)$ \\
\hline
Toronto (all) & 0.45 & 0.45 & 0.022 & 0.021 & [0.41, 0.49] & [0.41, 0.49] \\
Toronto Summer & 0.43 & 0.43 & 0.024 & 0.022 & [0.38, 0.48] & [0.39, 0.48] \\
Toronto Winter & 0.48 & 0.49 & 0.026 & 0.025 & [0.43, 0.53] & [0.44, 0.54] \\
Toronto Spring & 0.31 & 0.37 & 0.020 & 0.019 & [0.27, 0.35] & [0.33, 0.41] \\
Toronto Fall   & 0.41 & 0.42 & 0.039 & 0.033 & [0.33, 0.49] & [0.36, 0.48] \\
\hline
Chicago (all)  & 0.18 & 0.16 & 0.038 & 0.033 & [0.11, 0.25] & [0.10, 0.23] \\
Chicago Summer & 1.46 & 1.71 & 0.032 & 0.028 & [1.40, 1.52] & [1.66, 1.76] \\
Chicago Winter & 0.23 & 0.22 & 0.029 & 0.031 & [0.17, 0.29] & [0.16, 0.28] \\
Chicago Spring & 0.66 & 0.65 & 0.025 & 0.027 & [0.61, 0.71] & [0.60, 0.70] \\
Chicago Fall   & 0.03 & 0.01 & 0.025 & 0.027 & [-0.02, 0.08] & [-0.04, 0.06] \\
\hline
\end{tabular}
\end{table}

\subsection{Seasonal heterogeneity}

MLE results in Table~\ref{tab:gamma_parameters} show $\hat\alpha$ and
$\hat\beta$ differ markedly by season, especially in winter, motivating a
season-specific CNN fit. The network therefore uses only December data when
pricing the December contracts studied in Section~5.

\medskip
With $(\alpha,\beta)$ in hand—either from MLE or CNN—we move next to analytic
and Monte-Carlo pricing of precipitation strangle contracts.

\section{Weather-derivative pricing}

Most weather derivative contracts are based on accumulated temperatures or precipitations (CAT), heating-degrees-days (HDD) or cooling-degrees-days (CDD) over certain period $[0, \tau]$ containing $n$ days. A Pacific Rim index considers the average value of the climate underlying variable. The later is defined, respectively for temperature and precipitation, as:

\begin{eqnarray}\label{eq:prt}
 \xi_T&=&  \frac{1}{n}\sum_{k=1}^{n} T_k\\ 
 \xi_R&=&  \frac{1}{n}\sum_{k=1}^{n} R_k
\end{eqnarray}
In days with no precipitation $R_k=0$.\\
The payoff for these contracts are represented by a \textit{strangle contract}, i.e. a European long put and long call with different strikes. In this formula $d_1$ and $d_2$ represent the dollar amount paid per degree above or below the strike prices $K_1$ and $K_2$ respectively. This comes to \$20 (\$36 CAD) for contracts in  Fahrenheit degrees, and \$36 for contracts in Celsius degrees. Similar interpretation stands for $d_3, d_4, K_3$ and $K_4$. 

\begin{eqnarray*}\label{eq:payoff}
h(\xi_T) &=& d_1(\xi_T-K_1)_++d_2(K_2-\xi_T)_+ , d_1>0,d_2>0, K_1>K_2>0  \\
h(\xi_R) &=& d_3(\xi_R-K_3)_++d_4(K_4-\xi_R)_+ , d_3>0, d_4>0,,K_3>K_4>0 
\end{eqnarray*}
The price of WD contracts based on temperatures and precipitation are then given by:
\begin{eqnarray} 
p_T &=& e^{-r \tau} E[h(\xi_T)]= d_1 I_1 + d_2 I_2 \\ \label{eq:pricerain}
p_R &=& e^{-r \tau} E[h(\xi_R)]= d_3 I_3+d_4 I_4  
\end{eqnarray}
where
\begin{eqnarray} 
I_1 &=& \int_{\mathbb{R}} (x-K_1)_+ \; f_{\xi_{T}}(x,\theta)\;dx,\; 
 I_2 = \int_{\mathbb{R}} (K_2-x)_+ f_{\xi_{T}}(x,\theta) \;dx\\
 I_3 &=& \int_{\mathbb{R}}  (x-K_3)_+ \;f_{\xi_{R}}(x,\theta) \;dx, \;
 I_4 = \int_{\mathbb{R}}   (K_4-x)_+ f_{\xi_{R}}(x,\theta) \;dx  
 \end{eqnarray}
and $f_{\xi_{T}}(x,\theta)$ and $f_{\xi_{R}}(x,\theta)$ are the respective p.d.f.'s  of the Pacific Rim indices for temperatures and precipitation.\\
As pricing reference we use the Historic Burn approach (HBA), see \cite{Achilleas}. Despite of its inaccuracies HBA remains rather popular between practitioners. It consists in looking at past payoffs of the contract and compute their mean value conveniently discounted. A linear trend might be removed form the data to account for non-stationarity due to local change in urbanization or global warming. A premium is added to account for the risk of writing the contract, as much as 20\%-25\% of its standard deviation in practice. In summary the price using HBA can be written as:

\begin{equation*}
  p_{HBA}=e^{-rT}\left( \frac{1}{m} \sum_{k=1}^m h^{(k)}(\xi)+ 0.25 std( h(\xi)) \right)
\end{equation*}
where $h^{(k)}$ is the  value of the payoff on the k-th year, $std( h(\xi))$ is its standard deviation and $m$ is the number of years back in the analysis. Here $\xi$ is the Pacific Rim index for temperature or precipitation.\\
 The estimated payoff for the temperature contract is calculated based on the forecasts of the Harmonic/ARMA(2,3) and the neural network model. Alternatively,  a forecast for the Pacific Rim index can be estimated through Monte Carlo simulations of a gamma distribution that estimates precipitation for precipitation days.\\
 For the numerical study we carry on,  the strike prices are selected as the percentiles of temperature and precipitation. The level of the percentile ranges on the interval $[50,99]$.
 
\subsection{Pricing Toronto and Chicago temperatures}
\begin{figure}[hbt!]
\begin{center}
\subfigure[]{
\resizebox*{7cm}{!}{\includegraphics{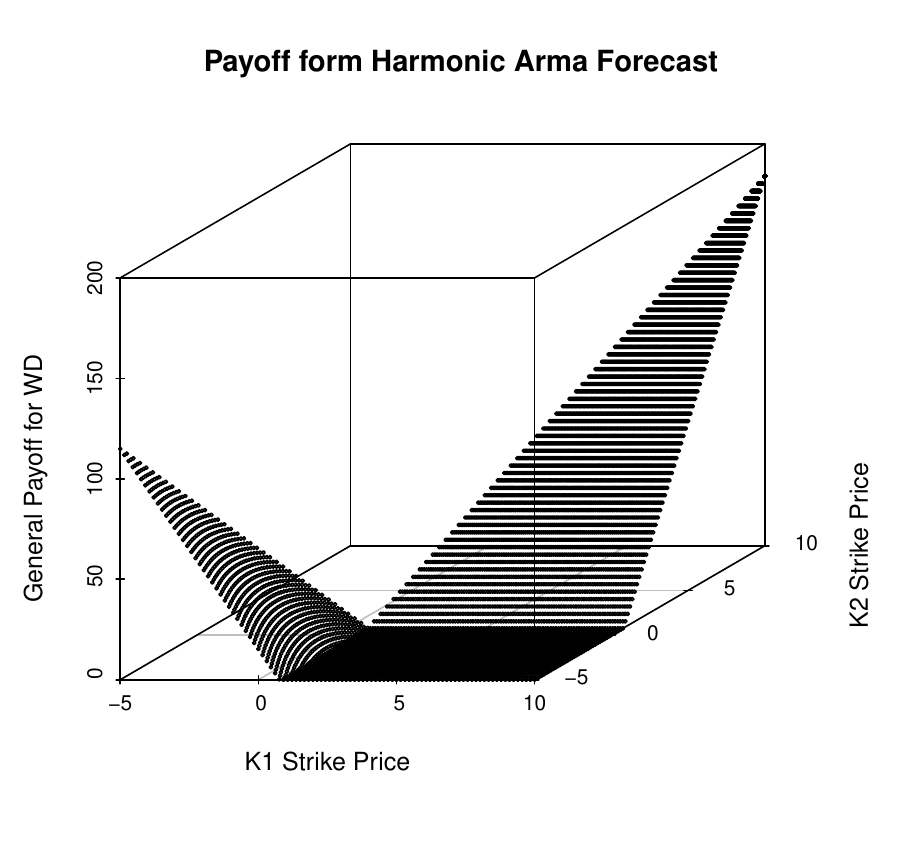}}}
\subfigure[]{
\resizebox*{7cm}{!}{\includegraphics{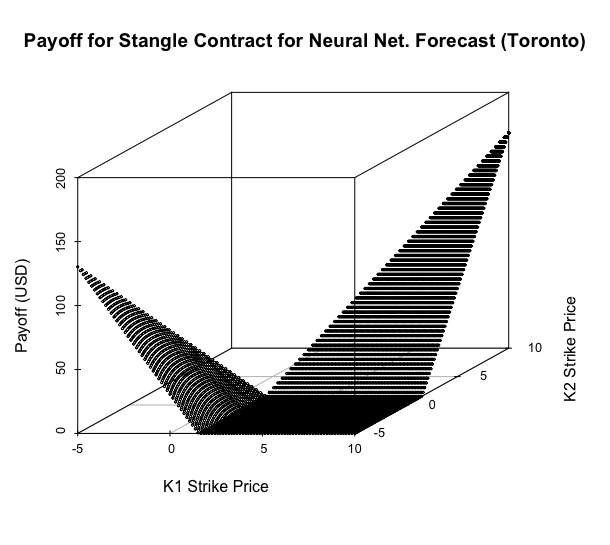}}}
 \caption{Payoff from harmonic/ARMA forecast and neural network for a European long put and a long call with different strikes (strangle) with different strike prices.}\label{fig:T_TO_payoff_scatter}
\end{center}
\end{figure}

In Table \ref{tab:TO_primsvals} we can see the forecast for the Pacific Rim index for Toronto and Chicago in December 2023. The first line represents the  value for the Pacific Rim index across the data time expand, while the second line shows the 40 year average of the index only for December from 1982 to 2022. The next three following lines represents the index forecast under the harmonic regression, the harmonic regression/ARMA and the one using neural networks. \\
The results indicate that neural networks are more capable at predicting daily average temperature when the forecast window is one month. Using the forecasts from the Harmonic/ARMA model and the neural network model we calculate the the estimated payoff for a strangle contract for different strike prices. 
 Confidence intervals in the forecasts for December for the harmonic/ARMA and neural network model can be seen in figures \ref{fig:hrm_arma_fore} and \ref{fig:T_TO_NN_fore}. We observe that the payoff of the WD strangle contract is larger for more extreme weather forecasts. The largest estimated payouts occur when we expect much lower temperatures, or much higher temperatures. This is why the values in the further bounds of the confidence intervals of  the weather forecast correspond to higher estimated payoffs. However, the payoffs are less likely to be observed, as they do not represent the most likely outcome.\\
 Having both a call and put option allows for a payoff when temperatures are either higher or lower. This makes a strangle contract a favorable contract to hold when temperatures could be much higher or lower than expected.\\
 Because a strangle contract can provide a payoff when temperatures are much higher or lower this contract could be especially useful in addressing economic issues related to climate change as temperatures become more volatile and less predictable. Figures \ref{fig:T_TO_payoff_scatter}(a) and \ref{fig:T_TO_payoff_scatter}(b) show a three dimensional scatter plot of the payoffs for the same contract when for different strike prices. Here the strike price for a call $(K_1)$ is unrelated to the strike price for the put $(K_2)$ with the exception that $K_1 > K_2$. 

\begin{table}
\centering
\begin{tabular}{|c|c|c|c}
\hline
Method &    PR Toronto &  PR Chicago &         \\ \hline
Actual value of PR   & 3.304 &  1.83 \\
40 Year Average  & -0.1634146&   -3.217561   \\
Harmonic Regression &   0.8564685& -2.477111 \\
Harmonic/ARMA model &   0.7683072 & -2.17967 \\
Neural Network model & 1.541371 &  -1.137333 \\
\hline
\end{tabular}
\caption{\label{tab:TO_primsvals} Estimates for the Pacific Rim index for Toronto and Chicago December 2023 using different methods}
\end{table}

\begin{figure}
    \centering
    \includegraphics[width=1\textwidth]{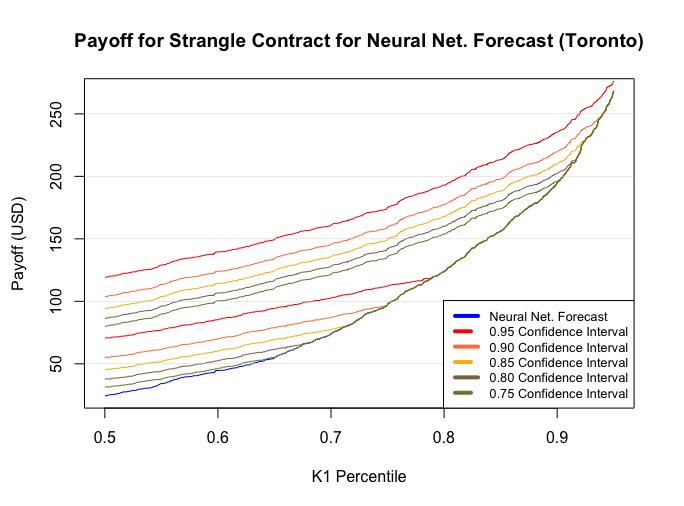}
    \caption{\label{fig:T_TO_NN_payoff_2D} Payoff from neural network forecast from European long put and a long call with different strikes (strangle) where the strike price is a percentile of December temperatures.}
\end{figure}

\begin{figure}[hbt!]
\begin{center}
\subfigure[]{
\resizebox*{7cm}{!}{\includegraphics{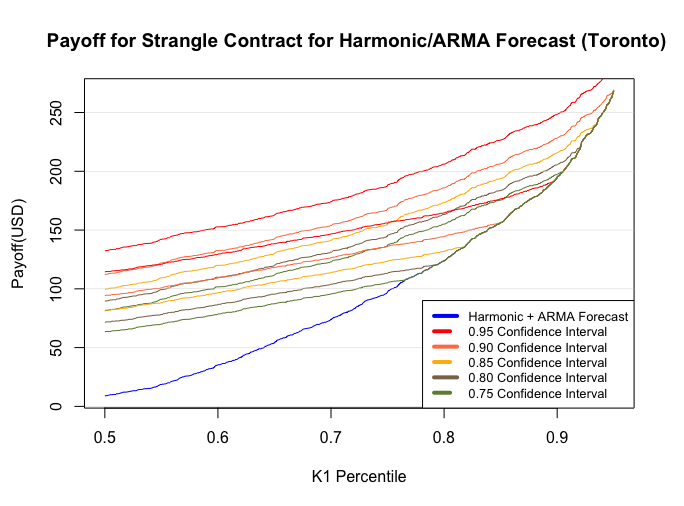}}}
\subfigure[]{
\resizebox*{7cm}{!}{\includegraphics{T_TO_NN_payoff_2D.png}}}
 \caption{ Price from harmonic/ARMA forecast from European long put and a long call with different strikes (strangle) where the strike price is a percentile of December temperatures(left). Price from neural network forecast(right) }\label{fig:harmonic-arma-marco-hening-2}
\end{center}
\end{figure}

\begin{figure}[hbt!]
\begin{center}
\subfigure[]{
\resizebox*{7cm}{!}{\includegraphics{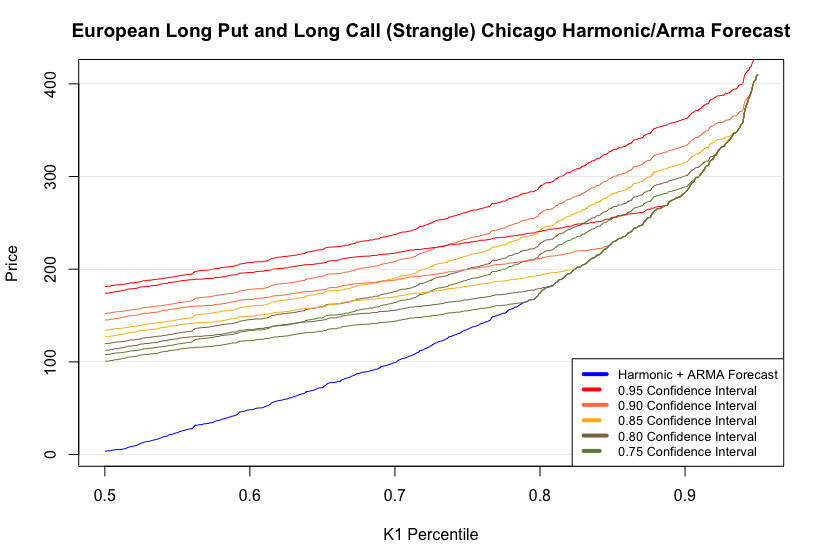}}}
\label{fig:harmonic-arma-marco-hening}
\subfigure[]{
\resizebox*{7cm}{!}{\includegraphics{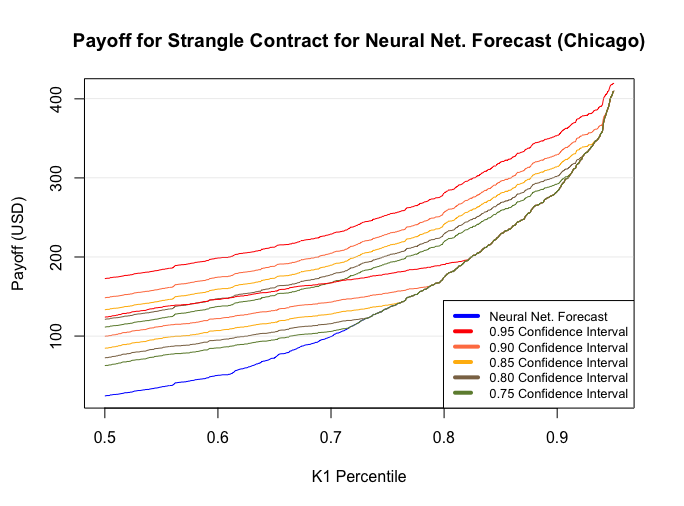}}}
 \caption{ Price from harmonic/ARMA forecast from European long put and a long call with different strikes (strangle) where the strike price is a percentile of December temperatures(left). Price from neural network forecast(right) }\label{fig:harmonic-arma-marco-hening-2b}
\end{center}
\end{figure}

\begin{figure}[hbt!]
\begin{center}
\subfigure[]{
\resizebox*{7cm}{!}{\includegraphics{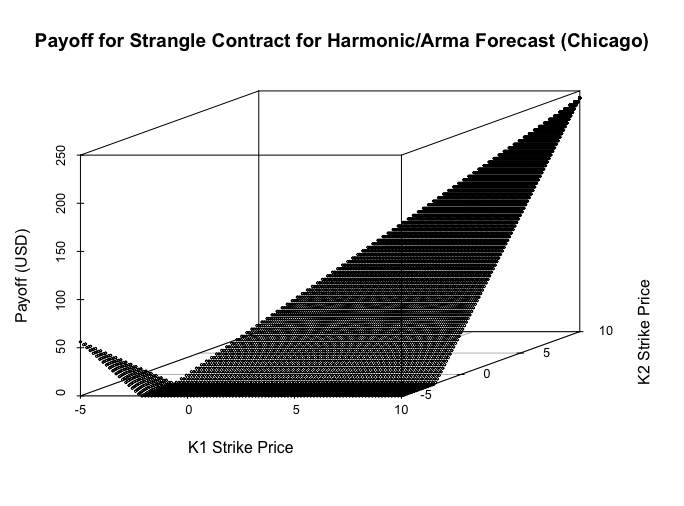}}}
\subfigure[]{
\resizebox*{7cm}{!}{\includegraphics{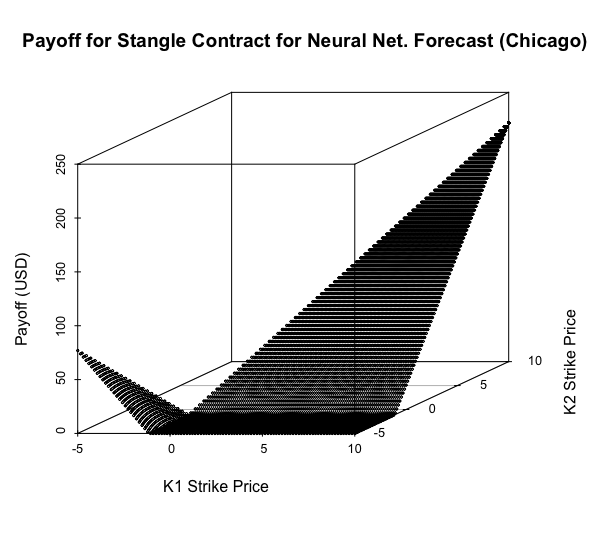}}}
 \caption{Payoff from harmonic/ARMA forecast from European long put and a long call with different strikes (strangle) with different strike prices(left) and payoff from neural network forecast9right)}\label{fig:T_CH_hrmarma_payoff_scatter}
\end{center}
\end{figure}

 A scatter plot of the the Pacific Rim for Chicago from 1981-2023 also shows the index is increasing. Figures \ref{fig:harmonic-arma-marco-hening}(a) and \ref{fig:harmonic-arma-marco-hening-2}(b) show the estimated payoff for a strangle based on the forecast of the Harmonic/ARMA and neural network model. In these figures the value $K_1$ is the strike price  for the call option given in terms of percentiles of the mean temperature for the month of December.  We can see that when the strike prices are chosen as percentiles of historical temperatures as explained in the introduction to this section we  get a non-linear increase in the payoff of the contract. In figures \ref{fig:T_CH_hrmarma_payoff_scatter}(a) and \ref{fig:T_CH_hrmarma_payoff_scatter}(b) where we can see the payoff for our harmonic/ARMA model and neural network model when we pick unrelated strike prices we see that the payoff for our contract increases linearly.

\subsection{Pricing Toronto and Chicago precipitation contracts}
\begin{figure}[hbt!]
\begin{center}
\subfigure[]{
\resizebox*{7cm}{!}{\includegraphics{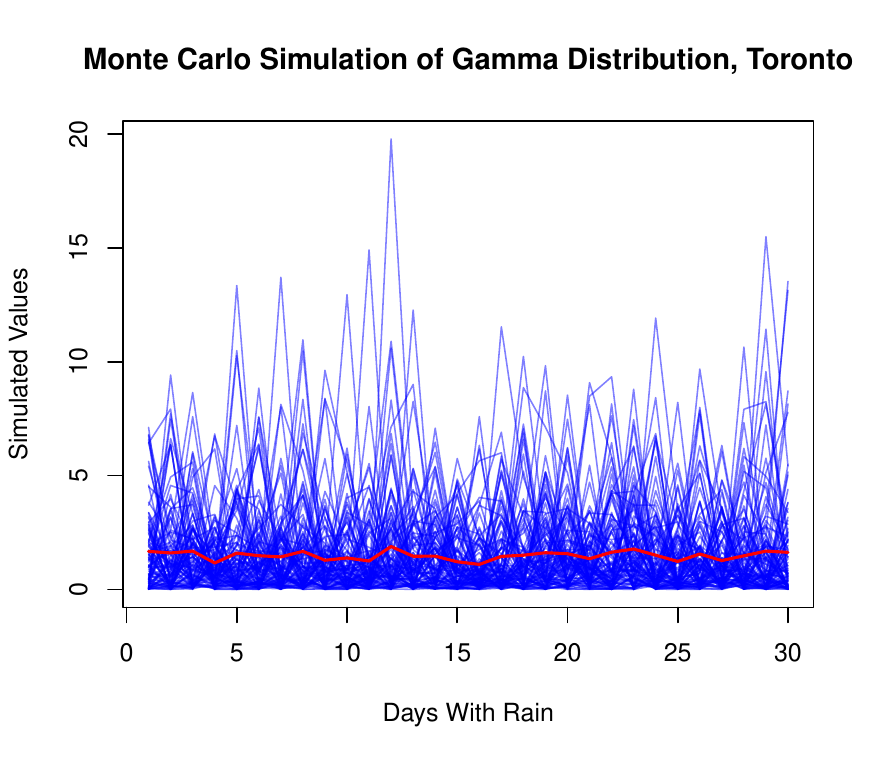}}}
\subfigure[]{
\resizebox*{7cm}{!}{\includegraphics{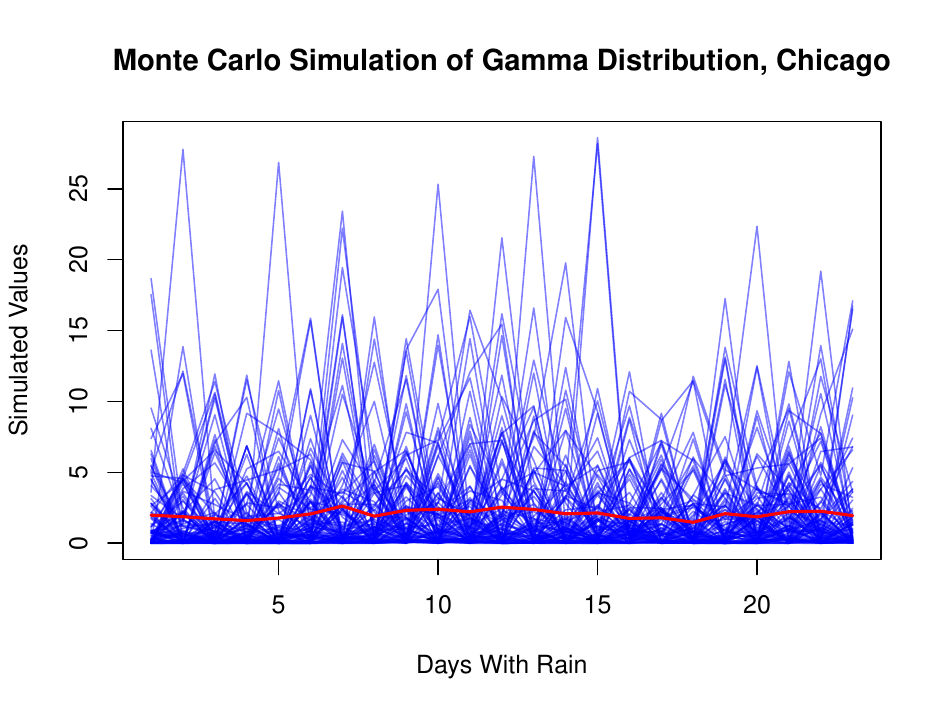}}}
 \caption{Monte Carlo simulation for rainy days in Toronto ($\alpha = 0.580, \beta = 0.377$) and Chicago ($\alpha = 0.354, \beta = 0.166)$}\label{fig:TO_monte}
\end{center}
\end{figure}

We show prices of WD contracts based on precipitation for the month of December. The cumulated precipitation follows equation \ref{eq:prt} with a Gamma distribution for the daily amounts. The parameters of the Gamma distribution have been estimated in section 4 using a CNN and a MLE approach. It allows to make a forecast for daily cumulative precipitation for December 2023 by computing the integrals $I_3$ and $I_4$ in equation (\ref{eq:pricerain}).\\
By mean of  the estimated values $\alpha$ and $\beta$ also  we can perform a Monte Carlo simulation for precipitations in Toronto and Chicago. The  no raining days  are modeled through a Poisson Process with parameter $\lambda >0$. We estimated the later by taking the average number of days it rains in December each year, from 1981-2023.  In total, an estimate of the DTP for December is obtained after 1000 simulations.\\
Graphs showing the simulated values of precipitation for Toronto and Chicago can be seen in figures \ref{fig:TO_monte}(a) and \ref{fig:TO_monte}(b) respectively. Each blue line in this graph represents one simulation for December, and the red line represents the mean of all the simulated values. If we take the mean value of this red line this would provide us with an estimate for the Pacific Rim index. 

The price of the WD contract can be obtained by directly computing integrals $I_1, I_2, I_3$ and $I_4$. Notice that the sum of independent Gamma distributed random variables, independent and equally distributed is also  Gamma distributed with parameters $n \alpha$ and $\beta$. Hence the p.d.f. of the Pacific Rim index is:
\begin{equation}
f_{\xi} = nR(n\alpha, \beta, nx) = n\frac{\beta^{n\alpha}} {\Gamma(n\alpha)}(nx)^{n\alpha-1} e^{-n\beta x}
\end{equation}

On the other hand, after elementary transformations we have:
\begin{eqnarray}
    I_3&=& \int_{-\infty}^{\infty} (x-K_3)_+f_{\xi} \, dx = \int_{K_1}^{\infty}xf_{\xi}\, dx - K_3\int_{K_1}^{\infty}f_{\xi} \, dx \\
    &=& \frac{1}{n\beta\Gamma(n\alpha)}\int_{t=n\beta K_3}^{\infty}t^{n\alpha}e^{-t}\, dt = \frac{\Gamma(n\alpha+1, n\beta K_3)}{n\beta \Gamma(n\alpha)}
\end{eqnarray}

Similarly:

\begin{equation}
  I_4 = \int_{-\infty}^{\infty} (k_2-x)_+f_{\xi} \, dx = -\frac{\gamma(n\alpha+1, n\beta k_2)}{n\beta \Gamma(n\alpha)} + k_2 \frac{\gamma(n\alpha, n\beta k_2)}{\Gamma(n\alpha)}
\end{equation}

By using the prices  for a call and put option we can  compute a pricing formula for a strangle contract, as in  equation \ref{eq:precip_payoff}.\\
Thus, using the upper-incomplete, lower-incomplete, and complete gamma function we can easily calculate the estimated payoff of a call, put, or strangle contract where precipitation is modeled using a compound Poisson process with the gamma distribution.\\
Gathering both terms $I_3$ and $I_4$ and replacing in the price formula (\ref{eq:pricerain}) we have the final price given by:

\begin{align}\label{eq:precip_payoff}
h(\xi_T) &= d_1(\xi_T-K_1)_+ + d_2(K_2-\xi_T)_+ \nonumber \\
&= d_1 \left(  \frac{\Gamma(n\alpha+1, n\beta K_3)}{n \beta \Gamma(n\alpha)} -
K_3 \frac{\Gamma(n\alpha, n\beta K_3)}{\Gamma(n\alpha)} \right) \nonumber \\
&\quad + d_2\left( -\frac{\gamma(n\alpha+1, n\beta k_2)}{n\beta \Gamma(n\alpha)} + k_2 \frac{\gamma(n\alpha, n\beta k_2)}{\Gamma(n\alpha)}\right)
\end{align}

\begin{figure}[htb!]
\begin{center}
\subfigure[]{
\resizebox*{7cm}{!}{\includegraphics{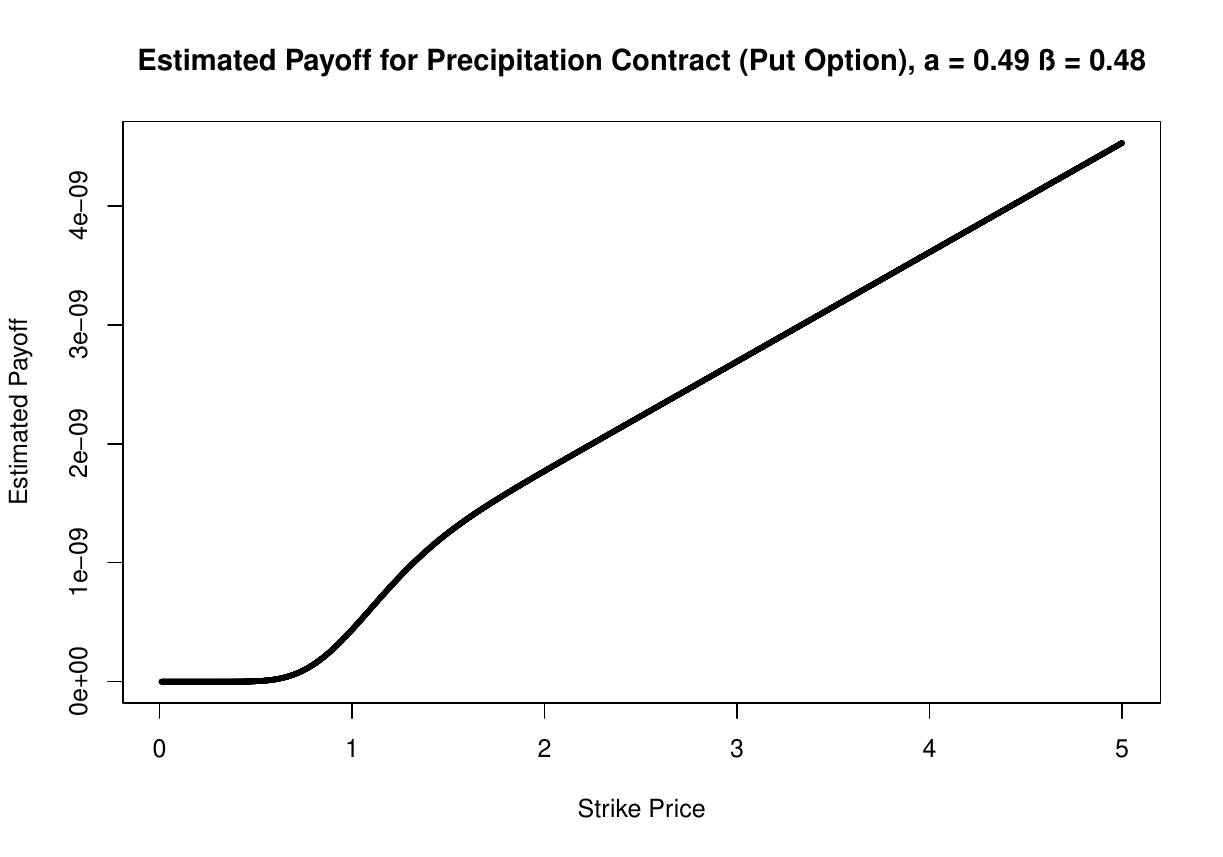}}}
\subfigure[]{
\resizebox*{7cm}{!}{\includegraphics{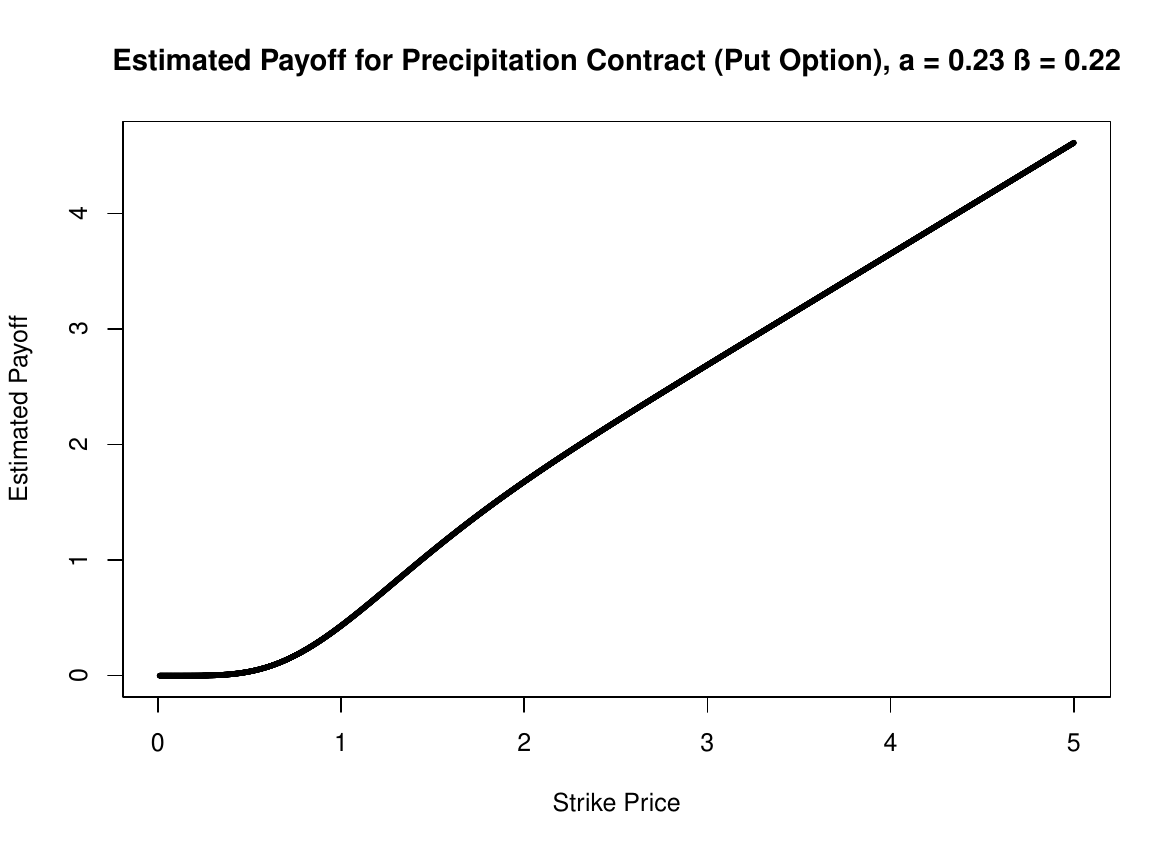}}}
 \caption{Price of the precipitation contract for Toronto(left) and Chicago(right) computed from the estimates of $\alpha$ and $\beta$ using the CNN.}\label{to_cnn_payoff}
\end{center}
\end{figure}

In figures \ref{to_cnn_payoff}(a) and \ref{to_cnn_payoff}(b) we can see the price of the part of the contract defined by the call option as function of the strike price for both cities. In both figures the $\alpha$ and $\beta$ values are the hyper-parameter values produced from the convolutional neural network in Section 4.2. We chose to use the estimates from the CNN because this method does not assume i.i.d. conditions, which is not an assumption supported by empirical evidence.

\section{Conclusions}

Using the 1981–2023 satellite-derived NASA POWER data set, we compared
traditional time-series techniques with neural-network methods for pricing
weather-derivative contracts on daily temperature and precipitation in Toronto
and Chicago.  Several clear messages emerge.

\medskip
\noindent\textbf{Temperature contracts.}
A feed-forward neural network produced a marked reduction in one-month-ahead
mean-squared error relative to the harmonic-regression/ARMA(2,3) benchmark.
That improvement is economically meaningful: fair strangle prices computed
from the neural forecasts differ noticeably from those based on the
time-series model and from those implied by the Historic Burn Approach,
especially for strikes deep in or out of the money.

\begin{itemize}
  \item \textbf{Toronto (Dec 2023).}
        \begin{itemize}
          \item Harmonic/ARMA reduces the climatological absolute error from
                $3.45^{\circ}\text{C}$ to $2.54^{\circ}\text{C}$, a 26 \% improvement.
          \item The neural network lowers it further to $1.76^{\circ}\text{C}$,
                yielding a 49 \% improvement over climatology and an additional
                23 percentage-point gain on the time-series model.
        \end{itemize}

  \item \textbf{Chicago (Dec 2023).}
        \begin{itemize}
          \item Harmonic/ARMA cuts the error from
                $5.03^{\circ}\text{C}$ to $4.01^{\circ}\text{C}$, a 20 \% improvement.
          \item The neural network achieves $2.97^{\circ}\text{C}$,
                corresponding to a 41 \% improvement over climatology and a further
                21 percentage-point advantage over Harmonic/ARMA.
        \end{itemize}
\end{itemize}

\medskip
\noindent\textbf{Precipitation contracts.}
Daily rainfall was modelled with a compound Poisson–Gamma process.
Gamma shape and scale were estimated by (i) maximum likelihood and
(ii) a convolutional neural network (CNN) trained on synthetically generated rainfall sequences. While the CNN’s parameter estimates are less precise, they adapt $(\alpha, \beta)$ to seasonal regimes—without modeling serial correlation—and (together with the Gamma‑sum formula) yield a closed‑form strangle price. This approach still assumes days are i.i.d.\ $\Gamma(\hat{\alpha}, \hat{\beta})$ at valuation and replaces the full compound Poisson–Gamma model with its mean‑count approximation $\Gamma(n\alpha, \beta n)$, trading off Monte‑Carlo overhead for an analytically tractable density.

\smallskip
\noindent\textbf{Seasonal heterogeneity.}
Table~\ref{tab:gamma_parameters_CNN} shows that the Gamma shape and scale parameters
change noticeably with season—for example, Toronto’s
$\hat\alpha$ rises from 0.516 in summer to 0.580 in winter—so the CNN used to
price the December strangles is trained only on December rainfall records.

\smallskip
\noindent\textit{Estimator precision.}
The same tables reveal that the classical
MLE is markedly more precise than the CNN:
across the ten season–city pairs
the CNN’s standard errors are, on average,
$3.9\times$ larger for $\hat\alpha$ and $6.3\times$ larger for
$\hat\beta$.\footnote{For each row of
Tables~\ref{tab:gamma_parameters}--\ref{tab:gamma_parameters_CNN}
we form the ratio
$\text{SE}_{\mathrm{CNN}} / \text{SE}_{\mathrm{MLE}}$ and then take the
arithmetic mean over all rows; the quoted factors are the resulting
averages for shape and scale, respectively.}
Thus, while the neural estimator sacrifices precision, it compensates by
learning the inter-day dependence that the i.i.d.\ MLE cannot capture.

\medskip
\noindent\textbf{Precipitation contracts.}
Daily rainfall was modelled with a compound Poisson–Gamma process.
Gamma shape and scale were estimated by (i) maximum likelihood and
(ii) a convolutional neural network (CNN) trained on synthetically generated
rainfall sequences.  
Although the CNN’s parameter estimates are less precise,
its resulting contract valuations remain competitive and, crucially,
reflect inter-day dependence.
Thanks to the Gamma-sum property we also obtained a closed-form formula for
precipitation strangles, eliminating Monte-Carlo simulation.
Taken together with the season-aware fitting strategy above,
these results suggest that a modest loss of statistical precision can be an
acceptable price for modelling structure that matters economically.

\medskip
\noindent\textbf{Model robustness.}
The CNN training and validation losses plateaued early, indicating no
discernible over-fitting.  Both modelling tracks, however, under-predicted the
unusually warm December 2023—an El Niño year—suggesting that future work
should incorporate large-scale climate indices such as ENSO or NAO.

\medskip
\noindent\textbf{Practical implications and future work.}
Neural-network methods capture nonlinear climatic dynamics that elude
classical techniques, improving pricing accuracy at the cost of wider
parameter-uncertainty bands—a trade-off that appears acceptable in practice.
Extending the horizon beyond one month, adding climate-mode indicators, and
testing the framework in other regions are promising next steps.
Finally, the NASA POWER data proved to be a reliable, globally available
alternative to sparse station observations, and we recommend its broader use in weather-risk applications.

\clearpage
\bibliographystyle{plain}
\bibliography{Referencesweatherts}

\end{document}